\documentstyle[aps,psfig,tighten,floats]{revtex}
\begin{document}

\title{{\bf 
Barrett-Crane model from a Boulatov-Ooguri field theory 
over a homogeneous space}}
\author{{ 
Roberto De Pietri ${}^{a,1}$,
Laurent Freidel   ${}^{b,c,2}$,
Kirill Krasnov    ${}^{b,d,3}$,
Carlo Rovelli     ${}^{a,d,e,4}$
}\\[2mm]
${}^{a}${\it Centre de Physique Th\'eorique - CNRS, Case 
907, Luminy,
             F-13288 Marseille, France } \\
${}^{b}${\it Center for Gravitational Physics and Geometry,
             Penn State University, Pa 16802, USA } \\
${}^{c}${\it Laboratoire de Physique, Ecole Normale 
Sup\'erieure de Lyon 
             46, all\'ee d'Italie, 69364 Lyon Cedex 07, 
France} \\
${}^{d}${\it Institute for Theoretical Physics, UCSB, 
             Santa Barbara, Ca 93106, USA}\\
${}^{e}${\it Physics Department, University of Pittsburgh, 
             Pittsburgh, Pa 15260, USA}
\\[2mm]
${}^{1}$ depietri@cpt.univ-mrs.fr,
${}^{2}$ freidel@phys.psu.edu,
${}^{3}$ krasnov@phys.psu.edu,
${}^{4}$ rovelli@pitt.edu 
\vskip.2cm}
\date{\today}
\maketitle

\begin{abstract}
\\PACS: 04.60.-m,04.60.Nc,11.10.-z.
\\Keywords: Quantum Gravity, Matrix Models.
\\[2mm]

Boulatov and Ooguri have generalized the matrix models of 2d 
quantum gravity to 3d and 4d, in the form of field theories 
over group manifolds.  We show that the Barrett-Crane 
quantum gravity model arises naturally from a theory of this 
type, but restricted to the homogeneous space 
$S^3=SO(4)/SO(3)$, as a term in its Feynman expansion.  From 
such a perspective, 4d quantum spacetime emerges as a 
Feynman graph, in the manner of the 2d matrix models.  This 
formalism provides a precise meaning to the ``sum over 
triangulations'', which is presumably necessary for a 
physical interpretation of a spin foam model as a theory of 
gravity.  In addition, this formalism leads us to introduce 
a natural alternative model, which might have relevance for 
quantum gravity.

\end{abstract}

%%%%%%%%%%%%%%%%%%%%%%%%%%%%%%%%%%%%%%%%%%%%%%%%%%%%%%%%
%% %% S E C T I O N %%
%%%%%%%%%%%%%%%%%%%%%%%%%%%%%%%%%%%%%%%%%%%%%%%%%%%%%%%%

\section{Introduction}

The Barrett-Crane relativistic model \cite{Barrett:1998} is 
an intriguing proposal for addressing the problem of 
constructing a quantum theory of gravity.  The model is 
related to the covariant, or spin-foam, formulation 
\cite{Reisenberger:1997} of loop quantum gravity 
\cite{loop}, and has recently received much attention.  
Barrett and Crane have defined this model as a state sum 
defined over a fixed triangulation $\Delta$ of 4d spacetime.  
Here, we show that the $q\to 1$ limit of the Barrett-Crane 
model can be derived from the Feynman expansion of a field 
theory over four copies of the homogeneous space $S^3={\rm 
SO}(4)/{\rm SO}(3)$.\footnote{Barrett and Crane define two 
versions of their model.  We recover here their ``first 
version''.} More precisely, the Barrett-Crane model is 
defined as a sum over colorings $c$ (assignments of spins to 
faces) of the triangulation $\Delta$
\begin{equation}
Z_{BC}[\Delta] = \sum_{c} A_{BC}(\Delta,c),
\label{c1}
\end{equation}
where $A_{BC}(\Delta,c)$ is the Barrett-Crane amplitude of 
the colored triangulation $(\Delta,c)$.  In the Feynman 
expansion
\begin{equation}
Z=\int {\cal D}\phi\, e^{-S[\phi]} = \sum_{\Gamma} 
{\lambda^{v[\Gamma]}\over {\rm sym}[\Gamma]} \ \ Z[\Gamma]
\label{field}
\end{equation}
of the field theory we consider, the Feynman graphs $\Gamma$ 
turn out to have the structure of a 2-complex.  Here 
$S[\phi]$ is the action, which depends on a coupling 
constant $\lambda$, while $v[\Gamma]$ and ${\rm 
sym}[\Gamma]$ are the number of vertices and the order of 
symmetries of $\Gamma$.  Every triangulation $\Delta$ 
determines a 2-complex $\Gamma=\Gamma(\Delta)$: the 
2-skeleton of its dual cellular complex.  We show in this 
paper that the Barrett-Crane model over the triangulation 
$\Delta$ is precisely the amplitude of the Feynman graph 
$\Gamma(\Delta)$:
\begin{equation}
  Z_{BC}[\Delta] = Z[\Gamma(\Delta)].
\end{equation}
More remarkably, the single terms of the sum (\ref{c1}) 
match.  In fact, first of all, the space over which our 
field theory is defined is compact and the Feynman integrals 
are replaced by Feynman sums
\begin{equation}
Z[\Gamma]= \sum_{c} \ \ A(\Gamma,c).
\label{c}
\end{equation}
Next, we show that the discrete ``momenta'' $c$ in (\ref{c}) 
correspond precisely to the colorings $c$ of the 
triangulation in (\ref{c1}).  That is, they are spins 
attached to (the 2-cells of the 2-complex $\Gamma(\Delta)$ 
which are dual to) the faces of $\Delta$.  And, finally, we 
show that the Barrett-Crane amplitude is precisely equal to 
the Feynman amplitude of the corresponding ``colored'' 
2-complex $(\Gamma, c)$
\begin{equation}
 A_{BC}(\Delta,c) = A(\Gamma(\Delta),c).
\end{equation}

The Feynman expansion of field theory we consider generates 
many more terms than a Barrett-Crane model over a fixed 
triangulation.  The additional summation is precisely what 
we need in order to address the main difficulty in 
developing a physical theory of quantum gravity starting 
from the Barrett-Crane model (or from spin foam models in 
general \cite{Baez}).  In these models, indeed, the choice 
of a fixed triangulation breaks the full covariance of the 
theory and unphysically restricts the number of the degrees 
of freedom.  Therefore some sort of ``sum over all 
triangulations'' is needed on physical grounds.  The 
difficulty is to choose a precise characterization of the 
objects over which to sum, and to fix the relative weights.  
This is precisely done by the full Feynman expansion of our 
field theory.  Therefore, the field theory provides a 
precise implementation of the loose idea of ``summing over 
triangulations''.  In detail, one first notices that the 
Barrett-Crane amplitude $A_{BC}(\Delta,c)$ does not depend 
on the full (combinatorial) information contained in the 
triangulation $\Delta$, but only on a subset of this.  In 
fact, it depends only on the 2-complex $\Gamma(\Delta)$, its 
dual 2-skeleton.  Thus, the Barrett-Crane state sum 
(\ref{c1}) is actually a sum over all colorings of a fixed 
2-complex.  The expansion (\ref{field}) extends the 
Barrett-Crane state sum (\ref{c1}) to a sum over colored 
2-complexes, and fixes the relative weight of these.  Thus, 
what emerges in the Feynman expansion is not a sum over 
actual {\em triangulations}, as one might have naively 
expected, but rather a sum over certain different objects: 
colored 2-complexes $s=(\Gamma,c)$
\begin{equation}
Z = \sum_{s} A(s).
\label{sf}
\end{equation}

Now, colored 2-complexes $s$ are precisely {\em spin foams}.  
Spin foams were derived (under the name branched colored 
surfaces) in describing the dynamics of loop quantum gravity 
\cite{Reisenberger:1997}.  Indeed, the covariant formulation 
of loop quantum gravity yields a partition function which is 
precisely a sum over spin foams of the form (\ref{sf}).  
Intuitively, a spin foam describes the time evolution of the 
spin networks, the states of loop quantum gravity 
\cite{spinnetwork}.  The idea that quantum spacetime could 
be described in terms of objects of this kind has been 
proposed earlier, in particular by Baez \cite{Baez:1994} and 
Reisenberger \cite{Reisenberger} (see also Iwasaki 
\cite{Iwasaki}).  In \cite{Baez}, Baez has defined and 
analyzed the general notion of {\em spin foam model\/} (and 
has introduced the term ``spin foam''); we refer to Baez' 
papers \cite{Baez} and \cite{Baez2} for an introduction and 
a discussion of these models.  See also 
\cite{othersn,altridue}.

The idea of obtaining a ``sum over triangulations'' from a 
Feynman expansion was successfully implemented some time 
ago, in the context of the matrix models of 2d dimensional 
quantum gravity, or ``zero dimensional string 
theory''\cite{2d}.  In that context, one views a 
triangulated spacetime as a term in a Feynman expansion of 
the matrix model.  Here, we essentially show that the same 
strategy works in 4d: the terms in the Barrett-Crane state 
sum can be loosely interpreted as a ``quantized spacetime 
geometry'', and we show here that these ``spacetime 
geometries'' are generated as Feynman graphs, as in the 2d 
quantum gravity models.  The convergence with the 2d matrix 
models is not accidental, since matrix models are at the 
root of one of the lines of development that lead to the 
Barrett-Crane theory.  Indeed, an extension of the matrix 
models to 3d was obtained by Boulatov in 
\cite{Boulatov:1992}.  Boulatov showed that the Feynman 
expansion of a certain field theory over three copies of 
${\rm SU}(2)$ generates triangulations, colorings and 
amplitudes of the Ponzano-Regge formulation of 3d quantum 
gravity \cite{Ponzano:1968} (and, taking the $q$ deformation 
of $SU(2)$, the amplitudes of the Turaev-Viro model 
\cite{Turaev-Viro}).  Ooguri has extended the Boulatov 
construction to 4d in \cite{Ooguri:1992b}.  The Ooguri 
theory is a field theory over 4 copies of ${\rm SU}(2)$.  
Its Feynman expansion determines a state sum for a 
triangulated 4d spacetime, in which the sum is over the 
irreducible representations of ${\rm SU}(2)$.  Replacing 
${\rm SU}(2)$ with the quantum group ${\rm SU}(2)_q$ ($q$ 
root of 1), yields a finite and well defined sum, the 
Ooguri-Crane-Yetter model \cite{Crane-Yetter}, which was 
shown by Crane, Kauffman and Yetter \cite{Crane} to be 
triangulation independent, and therefore to be an invariant 
of the 4-manifold.  The Ooguri-Crane-Yetter invariant can be 
shown to be the partition function of BF theory \cite{BF}.  
As euclidean general relativity can be seen as an ${\rm 
SO}(4)$ BF theory with an added constraint, see e.g.\ 
\cite{DF}, it was then natural to search for a model of 
quantum euclidean general relativity as a modification of 
the ${\rm SO}(4)$ Ooguri-Crane-Yetter state sum model.  The 
constraint that reduces BF theory to general relativity has 
an appealing geometrical interpretation.  Barrett and Crane 
realized that this constraint can be implemented within an 
${\rm SO}(4)$ Crane-Yetter state sum model by 
(appropriately) restricting the sum to simple 
representations only.\footnote{%
Irreducible representations of $SO(4)$ are labeled by two 
half integers (two spins) $(j',j"): j'+j''={\rm integer}$.  
A representation is simple if $j'=j":=j$.  Thus, simple 
representations are labeled by just one spin $j$.  Here, 
following group-theory literature conventions, we use the 
``color'' $N=2j$, instead of the spin $j$, for labeling the 
simple representation.} (See also \cite{Barbieri}.)\ Results 
in \cite{altridue,FW} support the idea that the 
Barrett-Crane model is indeed related to quantum general 
relativity.

Recently, it was pointed out that harmonic analysis on the 
homogeneous space $S^3= {\rm SO}(4)/{\rm SO}(3)$ yields 
precisely the simple representations of ${\rm SO}(4)$ 
\cite{Freidel:1999b}.  It was then tempting to conjecture 
that by suitably restricting the 4d Boulatov-Ooguri field 
theory to this homogeneous space one could get (the $q\to 1$ 
limit of) the Barrett-Crane model.  Here we show that this 
is indeed the case, with an additional bonus: while for the 
Ponzano-Regge and for the Ooguri-Crane-Yetter models the 
field theory generates a redundant sum over terms --all 
equal to each other by triangulation independence--, in the 
Barrett-Crane case the additional summation cures the 
breaking of covariance introduced with the choice of a 
triangulation, as we have discussed above.  This, in a 
certain sense, closes a circle, bringing the recent 
developments back to the original matrix model idea that 
spacetime can be viewed as a Feynman graph of a quantum 
theory.  The interpretation vividly emphasizes the 
background independence of these formulations of quantum 
gravity.

Finally, the formalism we develop, naturally suggest another 
model, different from Barrett-Crane one.  Indeed, in the 
field theory we consider, ${\rm SO}(4)$ invariance can be 
imposed in two ways: by the left or the right action of 
${\rm SO}(4)$ on ${\rm SO}(4)/{\rm SO}(3)$.  These turn out 
to be inequivalent.  The second alternative (Case B, below) 
yields the Barrett-Crane model (first version).  The first 
alternative (Case A, below), is far more natural from the 
group theoretical point of view.  It yields a different 
model, somewhat closer to, but distinct from the second 
version of the Barrett-Crane model.  We argue in the 
conclusion that this other model has interesting properties.  
In particular, it represents another solution to the problem 
of quantizing the constraints that transform BF theory into 
general relativity.

In section \ref{sec:def} we define our field theory, and 
state our main results in detail.  In section \ref{sec:mode} 
we find the mode expansion of the field.  Section 
\ref{sec:feyn} studies the Feynman expansion for the two 
theories.  In section \ref{sec:triang} we discuss the 
relation between the graphs generated by the Feynman 
expansion and the triangulations.  We also discuss some 
further variants of the model, in which orientation is taken 
into account.  In Section \ref{sec:disc} we conclude with 
some general comments.  In the two Appendices, we recall 
some elements of ${\rm SO}(N)$ representation theory.

% Since the graphical notation we used in previous works has
% proven hard to understand to many, we use here a
% conventional fully tensorial notation, at the price of a
% certain abundance of indices.

%%%%%%%%%%%%%%%%%%%%%%%%%%%%%%%%%%%%%%%%%%%%%%%%%%%%%%%%%%%%
%% S E C T I O N %%
%%%%%%%%%%%%%%%%%%%%%%%%%%%%%%%%%%%%%%%%%%%%%%%%%%%%%%%%%%%%

\section{Definition of the models and main result}
\label{sec:def}

We consider the group $G={\rm SO}(4)$ and we fix an ${\rm 
SO}(3)$ subgroup $H$ of $G$.  In particular, we may consider 
${\rm SO}(4)$ in its fundamental representation and choose a 
vector $v^0$ in the representation space; we then let $H$ be 
the ${\rm SO}(3)$ subgroup which leaves $v^0$ invariant.  
Right multiplication by an element of $H$ defines an 
equivalence relation in $G$: $g\sim g'$ if there is $h\in H$ 
such that $gh=g'$.  The space of the equivalence classes, 
${\rm SO}(4)/{\rm SO}(3)$, is diffeomorphic to the 3-sphere 
$S^3$.  We denote the elements of $S^3$ as $x$.  Note that 
the invariant normalized measures $dx$, $dh$ and $dg$ on 
$S^3$, ${\rm SO}(3)$ and ${\rm SO}(4)$ respectively are 
related as $dg=dx~dh$.

\begin{figure}
\centerline{{\psfig{figure=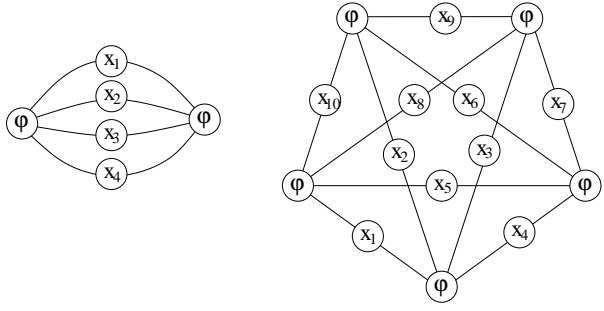,height=4cm}}}
\bigskip \caption{The structure of the kinetic and potential 
terms in the action.}
\label{fig:action}
\end{figure}

We consider a real field $\tilde\phi(g_1,g_2,g_3,g_4)$ over the 
cartesian product of four copies of $G={\rm SO}(4)$.  We 
take $\tilde\phi$ square integrable with respect to each argument.  
We require that $\tilde\phi$ is constant along the equivalence 
classes.
\begin{equation} \label{hinv}
        \tilde\phi(g_1,g_2,g_3,g_4) = \tilde\phi(g_1h_1,g_2h_2 ,g_3h_3 , 
        g_4h_4), \qquad\qquad (\forall \ h_1,h_2,h_3,h_4 \in H);
\end{equation}
so that $\tilde\phi$ is in fact a function over the homogeneous 
space $(S^3)^4$, which we can write as 
$\tilde\phi(x_1,x_2,x_3,x_4)$.  We will employ both notations.  We 
require $\tilde\phi$ to be invariant under any permutation 
$\sigma$ of its four arguments
\begin{equation} \label{perm}
\tilde\phi(g_{1},g_{2},g_{3},g_{4})\ =\ 
\tilde\phi(g_{\sigma(1)},g_{\sigma(2)}, 
g_{\sigma(3)},g_{\sigma(4)}).
\end{equation}
(In Section \ref{sec:triang}, we will consider alternative 
symmetry requirements as well.)  Finally, we project $\tilde\phi$ 
on its $SO(4)$ invariant part.  This can be done in two ways.
\begin{itemize}
\item {\bf Case A.} The natural action of ${\rm SO}(4)$ on 
the right coset $S^3={\rm SO}(4)/{\rm SO}(3)$ is the left 
action of the group.  Thus we define
\begin{equation} \label{linv}
        \phi(g_1,g_2,g_3,g_4) = 
        \int dg\ \tilde{\phi}(g g_1,g g_2,g g_3,g g_4) .
\label{CaseA}
\end{equation}
\item {\bf Case B.} Alternatively, we define
\begin{equation} \label{ginv}
        \phi(g_1,g_2,g_3,g_4) = 
        \int dg\ \tilde{\phi}(g_1g,g_2g,g_3g,g_4g) .
\label{CaseB}
\end{equation}
\end{itemize}
The two different invariance properties, (\ref{CaseA}) and 
(\ref{CaseB}) define two different theories.  In both cases, 
the dynamics is defined by the action
\begin{eqnarray}
S[\phi] &=& \frac{1}{2} \int \prod_{i=1}^{4} dg_i ~ 
\phi^{2}(g_1,g_2,g_3,g_4) \ + \frac{\lambda}{5!} \int 
\prod_{i=1}^{10} dx_i ~~\phi(g_1,g_2,g_3,g_4)\ 
\label{action} \\
\nonumber
&& \phi(g_4,g_5,g_6,g_7)\ \phi(g_7,g_3,g_8,g_9)\ 
\phi(g_9,g_6,g_2,g_{10}) \ \phi(g_{10},g_8,g_5,g_1).
\end{eqnarray}
The potential (fifth order) term has the structure of a 
4-simplex.  That is, if we represent each of the five fields 
in the product as a node with 4 legs --one for each $g_i$-- 
and connect pair of legs corresponding to the same argument, 
we obtain (the one-skeleton of) a 4-simplex, see Fig.  
\ref{fig:action}.  This completes the definition of our 
field theory.

\bigskip

In the following sections, we show that the Feynman 
expansion (\ref{field}) of this field theory is a sum over 
spin foams $s$ as in (\ref{sf}).  A spin foam $s=(\Gamma, 
c)$ is here a colored combinatorial 2-dimensional cellular 
complex, or colored 2-complex.  A 2-complex $\Gamma$ is 
formed by three abstract sets $\cal V, E, F$, whose elements 
are respectively denoted ``vertices'' $v$, ``edges'' $e$, 
and ``faces'' $f$, and boundary relations that assign two 
vertices to each edge, and a cycle of edges (a cyclic 
sequence of edges in which any two consecutive edges are 
distinct and share a vertex) to each face.  The number of 
edges that a vertex bounds (``adjacent'' to the vertex) is 
called the valence of the vertex, and the number of faces 
that an edge bounds (adjacent to the edge) is called the 
valence of the edge.  A coloring $c=(\{N_{f}\},\{i_{e}\})$ 
of the 2-complex $\Gamma$ is an assignment of an irreducible 
representations $N_{f}$ of some group $G$ to each face $f$ 
and of an ``intertwiner'' $i_{e}$ to each edge $e$.  An 
intertwiner $i_{e}$ is a vector in $K_{\vec N_{e}}$, the 
invariant subspace of the tensor product of the Hilbert 
spaces of the representations ${\vec N_{e}}=(N_{f_1}\ldots 
N_{f_n})$ assigned to the faces $f_{1\ldots f_{n}}$ adjacent 
to the edge $e$.  See the appendix for details on $K_{\vec 
N_{e}}$, and \cite{Baez,Baez2} for details on these models 
in general.

A generic spin foam model is defined by a sum over spin 
foams:
\begin{eqnarray}
        Z \ &=&\ \sum_{\Gamma} \ \ w(\Gamma)\ \ Z[\Gamma] \\
        Z[\Gamma] &=& \sum_{\{N_f\}, \{i_e\}} \ \prod_f A_f(N_f) 
        \ \prod_e A_e(\vec{N_e}, i_e) \ \prod_v 
        A_v(\vec{N_v},\vec{\imath}_v).
\end{eqnarray}
The first sum is over the 2-complexes and the second over 
the colorings.  $w(\Gamma)$ is a weight factor for each 
2-complex.  We have indicated by $\vec{N_e}$ the set of the 
colors of the faces adjacent to the edge $e$, and by 
$\vec{N_v}$ and $\vec{\imath}_v$ the set of the colors of 
the faces and the edges adjacent to the vertex $v$.  The 
factors $A_f$, $A_e$ and $A_v$ are called the amplitudes of 
the faces, edges and vertices.  Since every edge connects 
exactly two vertices, one can always absorb the amplitude of 
the edge $A_e(\vec{N_e},\vec{\imath}_e)$ into the definition 
of the vertex amplitude $A_v(\vec{N_v},\vec{\imath}_v)$.  
This is not possible for the amplitude of the face 
$A_f(N_f)$, since faces can be shared by an arbitrary number 
of vertices.

In this paper, we consider only 2-complexes $\Gamma$ with 
4-valent edges and 5-valent vertices (for any other 
$\Gamma$, we have $w(\Gamma)=0$).  Notice that the 
2-skeleton of the cellular complex dual to a triangulation 
$\Delta$ of a 4-manifold is a 2-complex of this kind.  
Indeed, vertices are 5-valent because they are dual to 
4-simplices, which are bounded by precisely five tetrahedra; 
and edges are 4-valent because they are dual to the 
tetrahedra, which are bounded by four triangles.

The Feynman expansion of our field theory yields a spin foam 
model.  The 2-complex weight factor $w(\Gamma)$ vanishes 
unless $\Gamma$ has 4-valent edges and 5-valent vertices 
only; in this case, it is given by $w(\Gamma) = 
\lambda^{n(\Gamma)} /sym(\Gamma)$.  Here $n(\Gamma)$ is the 
number of vertices and $sym(\Gamma)$ the order of the 
symmetries of the graph underlying the two-complex $\Gamma$, 
defined as in standard Feynman graph theory, see for 
instance \cite{Peskin}, page 93.  The amplitudes are as 
follows.
\begin{itemize}
\item {\bf Case A.} We color faces of by {\em simple\/} 
representation $N_f$ of $SO(4)$, and edges by intertwiners 
${i}_{e}$, forming an {\em orthonormal\/} basis in $K_{\vec 
N_{e}}$.  The amplitude of the face is $A_{f} =1$.  The 
amplitude of the edge is $A_e = 1$.  The vertex amplitude 
$A_v(\vec{N_v},\vec{\imath_v})$ is the $15$-$j$ Wigner 
symbol $A(\vec{N_v},\vec{\imath_v})$ constructed from the 
ten simple representations $\vec{N_v}$ and five intertwiners 
$i_v$.  The amplitude of a 2-complex is thus
\begin{equation}\label{case-a}
Z_A[\Gamma] = \sum_{\{N_f\}, \{i_e\}}\ \prod_v 
A(\vec{N_v},\vec{\imath_v}).
\end{equation}
\item {\bf Case B.} In this case, as in case A, faces are 
colored by simple representation $N_f$ of $SO(4)$.  Each 
edge, however, is colored by a single, fixed, intertwiner 
$i_{BC}$, the ({\em normalized}) Barrett-Crane intertwiner 
\cite{Barrett:1998}.  The amplitude of the faces is 
$A_{f}(N_f) ={\rm dim}_{N_{f}}$ --for simple 
representations, ${\rm dim}_N=(N+1)^2$.  The amplitude of 
the vertex $T_v(\vec{N_v},\vec{\imath_v})$ is the $15$-$j$ 
Wigner symbol constructed from the ten simple 
representations $\vec{N_v}$ and the five Barrett-Crane 
intertwiners $i_{BC}$.  Thus, the amplitude of a 2-complex 
is
\begin{equation}\label{case-b}
Z_B[\Gamma] = \sum_{\{N_f\}}\ \ \prod_f {\rm dim}_{N_f}\ \ 
\prod_v A(\vec{N_v},i_{BC}).
\label{B}
\end{equation}
This is precisely the Barrett-Crane amplitude for a 
triangulation $\Delta$ whose dual 2-skeleton is 
$\Gamma$.\footnote{Reference \cite{Barrett:1998} is a bit 
obscure concerning the amplitudes of the lower dimensional 
simplices of the triangulation, for which it refers to 
\cite{Crane-Yetter}, where a different normalization is 
used.  A careful analysis of the requirements put on these 
factors, and in particular of the requested invariance under 
change of basis in $K_{\vec N_{e}}$ (see 
\cite{Barrett:1998}, section IV), implies that the correct 
factors (using our normalization) are precisely as in 
(\ref{B}).}
\end{itemize}

%%%%%%%%%%%%%%%%%%%%%%%%%%%%%%%%%%%%%%%%%%%%%%%%%%%%%%%
%% S E C T I O N
%%%%%%%%%%%%%%%%%%%%%%%%%%%%%%%%%%%%%%%%%%%%%%%%%%%%%%%

\section{Mode expansion}
\label{sec:mode}

Consider a square integrable function $\phi(g)$ over 
$SO(4)$, invariant under the right action of $SO(3)$.  Using 
Peter-Weyl theorem, one can expand it in the matrix elements 
$D^{(\Lambda)}_{\alpha\beta}(g)$ of the irreducible 
representations $\Lambda$
\begin{equation}
\phi(g) = \sum_\Lambda \ \phi^{\Lambda}_{\alpha\beta} \ 
D^{(\Lambda)}_{\alpha\beta}(g)
\end{equation}
The indices $\alpha,\beta$ label basis vectors in the 
corresponding representation space, and the sum over 
repeated indices is understood.  In other words, we choose a 
basis in the representation space such that the metric in 
this basis is just the Kronecker delta, which is the 
standard choice in representation theory literature.  The 
requirement of invariance under the right ${\rm SO}(3)$ 
action can be written as
\begin{eqnarray}
\phi(g) &=& \int\limits_{SO(3)} dh \ \phi(gh).
\end{eqnarray}
Expanding this into the modes, we have
\begin{equation}
\phi(g) = \sum_\Lambda \ \phi^{\Lambda}_{\alpha\beta}\ 
D^{(\Lambda)}_{\alpha\beta}(g) = \int\limits_{SO(3)} dh\ 
\phi(gh) = \sum_\Lambda \ \int\limits_{SO(3)} dh\ 
\phi^{\Lambda}_{\alpha\beta}\ 
D^{(\Lambda)}_{\alpha\gamma}(g)\ 
D^{(\Lambda)}_{\gamma\beta}(h).
\end{equation}
The integral in the last term projects the $\beta$ index 
over a (normalized) $SO(3)$ invariant vector, which we 
denote by $w_\beta$.  As this exists only in simple 
representations (and is unique, see the appendix), the sum 
over representations $\Lambda$ reduces to a sum over the 
simple representations $N$ only, and thus
\begin{equation}
\phi(g) = \sum_N \ \phi^N_\alpha\ D^{(N)}_{\alpha\beta}(g)\ 
w_\beta .
\end{equation}
The quantities $H^{N}_\alpha(g)=D^{(N)}_{\alpha\beta}(g) 
w_\beta$ are invariant under the (right) action of $SO(3)$, 
and, thus, can be thought of as functions on the 
three-sphere: $H^{N}_\alpha(g)=H^{N}_\alpha(x)$.  In fact, 
they form an orthogonal basis for $S^3$ spherical harmonics.

Since $\phi$ is real, we have
\begin{equation}
\sum_N \ \overline{\phi^N_\alpha}\ 
\overline{D^{(N)}_{\alpha\beta}(g)} \ \overline{w_\beta} = 
\sum_N \ \phi^N_\alpha\ D^{(N)}_{\alpha\beta}(g) \ w_\beta .
\end{equation}
In the appendix, we show that the invariant vectors 
$w_\beta$ are real, and that the matrix elements can also be 
chosen to be real.  Thus, the reality condition simply 
requires $\phi^N_\alpha$ to be real.

Let us now consider the field $\tilde\phi(g_1,g_2,g_3,g_4)$.  From 
the observations above, it follows that the property 
(\ref{hinv}) implies that the field can be expanded as
\begin{equation} \label{genPHI}
\tilde\phi(g_1,\ldots,g_4) = \sum_{N_1\ldots N_4}\ \phi^{N_1\ldots 
N_4}_{\alpha_1\ldots\alpha_4}\ 
D^{(N_1)}_{\alpha_1\beta_1}(g_1) \ldots 
D^{(N_4)}_{\alpha_4\beta_4}(g_4)\ w_{\beta_1}\ldots 
w_{\beta_4}.
\end{equation}
As before, the sum over repeated internal indices is 
understood.  Finally, the symmetry of $\tilde\phi(g_{1}, \ldots , 
g_{4})$ under permutation of its four arguments implies that
\begin{equation} \label{permutation}
         \phi^{N_1\ldots N_4}_{\alpha_1\ldots\alpha_4} = 
         \phi^{N_\sigma(1)\ldots 
         N_\sigma(4)}_{\alpha_\sigma(1)\ldots\alpha_\sigma(4)},
\end{equation}
where $\sigma$ is any permutation of $\{1,2,3,4\}$.

Let us now find find the effect of the $SO(4)$ invariance 
property on the modes, and find the expression for the 
action in terms of the modes.  The analysis is different for 
the two cases A and B.

\bigskip \noindent{\bf Case A.} Substituting the mode expantion
(\ref{genPHI}) into the definition 
of $\phi$ 
\begin{eqnarray}
\phi(g_1\ldots g_4) &=& \int\limits_{SO(4)} dg \ \ \phi(g 
g_1,\ldots, g g_4),
\end{eqnarray}
and using formula (\ref{app:4-int}) for integral of the 
product of four group elements, one obtains
\begin{equation}
\phi(g_1,\ldots,g_4) = \sum_{N_1\ldots N_4} \sum_\Lambda 
\left(\phi^{N_1\ldots N_4}_{\alpha_1\ldots\alpha_4}\ 
C^{N_1\ldots N_4\,\Lambda}_{\alpha_1\ldots\alpha_4}\right) 
\left({C}^{N_1\ldots N_4\,\Lambda}_{\beta_1\ldots\beta_4} 
D^{(N_1)}_{\beta_1\gamma_1}(g_1) \ldots 
D^{(N_4)}_{\beta_4\gamma_4}(g_4)\ w_{\gamma_1}\ldots 
w_{\gamma_4}\ \right),
\end{equation}
where, notice, the sum over $\Lambda$ is over all --not 
necessarily simple-- representations.  Here $C^{N_1\ldots 
N_4\,\Lambda}_{\alpha_1\ldots\alpha_4}$ are four-valent 
intertwiners defined in the appendix.  It is clear from the 
above formula that the modes $\phi^{N_1\ldots 
N_4}_{\alpha_1\ldots\alpha_4}$ enter the mode decomposition 
only contracted with the intertwiner $C$.  Defining
\begin{equation} \label{def:phiA}
\phi_{\cal A}^{N_1\ldots N_4\,\Lambda}:= \phi^{N_1\ldots 
N_4}_{\alpha_1\ldots\alpha_4}\ C^{N_1\ldots 
N_4\,\Lambda}_{\alpha_1\ldots\alpha_4} \frac{1}{\sqrt{{\rm 
dim}_{N_1} \ldots {\rm dim}_{N_4}}},
\end{equation}
the mode expansion takes the form
\begin{equation}
\phi(g_1,\ldots,g_4) = \sum_{N_1\ldots N_4} \sum_\Lambda 
\phi_{\cal A}^{N_1\ldots N_4\,\Lambda} \sqrt{{\rm 
dim}_{N_1}\ldots {\rm dim}_{N_4}} \left({C}^{N_1\ldots 
N_4\,\Lambda}_{\beta_1\ldots\beta_4} 
D^{(N_1)}_{\beta_1\gamma_1}(g_1) \ldots 
D^{(N_4)}_{\beta_4\gamma_4}(g_4)\ w_{\gamma_1}\ldots 
w_{\gamma_4}\ \right).
\end{equation}
We now write the action in terms of the modes $\phi_{\cal 
A}$.  Using the result (\ref{app:2-int}) for the integral of 
the product of two matrix elements, the kinetic term becomes
\begin{equation}
\frac{1}{2} \int \prod_{i=1}^4 dg_i ~ 
~~\phi^2(g_1,\ldots,g_4) = \frac{1}{2} \sum_{N_1\ldots N_4} 
\sum_\Lambda \phi_{\cal A}^{N_1\ldots N_4\,\Lambda} 
\phi_{\cal A}^{ N_1\ldots N_4\, \Lambda}.
         \end{equation}
For the interaction term, we obtain
\begin{eqnarray}
&&{\lambda \over 5!} \int \prod_{i=1}^{10} dg_i 
~~\phi(g_1,g_2,g_3,g_4)\ \phi(g_4,g_5,g_6,g_7)\ 
\phi(g_7,g_3,g_8,g_9)\ \phi(g_9,g_6,g_2,g_{10}) \ 
\phi(g_{10},g_8,g_5,g_1) \\ \nonumber && = {\lambda \over 
5!} \sum_{N_1\ldots N_{10}} \sum_{\Lambda_1\ldots\Lambda_5} 
\phi_{\cal A}^{N_1 N_2 N_3 N_4\,\Lambda_1} \phi_{\cal 
A}^{N_4 N_5 N_6 N_7\,\Lambda_2} \phi_{\cal A}^{N_7 N_3 N_8 
N_9\,\Lambda_3} \phi_{\cal A}^{N_9 N_6 N_2 
N_{10}\,\Lambda_4} \phi_{\cal A}^{N_{10} N_8 N_5 
N_1\,\Lambda_5} ~~ {\cal 
A}_{N_1,\ldots,N_{10},\Lambda_1,\ldots,\Lambda_5}\ .
\end{eqnarray}
Here ${\cal 
A}_{N_1,\ldots,N_{10},\Lambda_1,\ldots,\Lambda_5}$ is a 
($SO(4)$ analog of the) $(15-j)$-symbol, defined as
\begin{equation}
{\cal A}_{N_1,\ldots,N_{10},\Lambda_1,\ldots,\Lambda_5} = \ 
{C}^{ N_1 N_2 N_3 
N_4\,\Lambda_1}_{\alpha_1\alpha_2\alpha_3\alpha_4} \ {C}^{ 
N_4 N_5 N_6 
N_7\,\Lambda_2}_{\alpha_4\alpha_5\alpha_6\alpha_7} \ {C}^{ 
N_7 N_3 N_8 
N_9\,\Lambda_3}_{\alpha_7\alpha_3\alpha_8\alpha_9} \ {C}^{ 
N_9 N_6 N_2 
N_{10}\,\Lambda_4}_{\alpha_9\alpha_6\alpha_2\alpha_{10}} \ 
{C}^{N_{10} N_8 N_5 
N_1\,\Lambda_5}_{\alpha_{10}\alpha_8\alpha_5\alpha_1},
\label{A}
\end{equation}
where summation over repeated internal indices is 
understood.  Notice that the indices are connected as in 
Figure 1, or Figure 3.  In (\ref{A}), the intertwiners are 
written in the basis (\ref{basis1}).  For a generic choice 
of basis elements $i_{e}^{(a)}$ in the Hilbert spaces 
$K_{\vec N_{e}}$
\begin{equation}
i_{e}{}^{ N_1 N_2 N_3 N_4}_{\alpha_1\alpha_2\alpha_3\alpha_4}{}^{(a)} = 
\sum_{\Lambda} \ M_{e}{}^{(a)}_{\Lambda}\ {C}^{ N_1 N_2 N_3 
N_4\,\Lambda}_{\alpha_1\alpha_2\alpha_3\alpha_4},
\end{equation}
we can write the vertex amplitude as
\begin{equation}
A(\vec N_{v},\vec i_{v}) = \ {i}^{ N_1 N_2 N_3 
N_4\,(a_1)}_{\alpha_1\alpha_2\alpha_3\alpha_4} \ {i}^{ N_4 
N_5 N_6 N_7\,(a_2)}_{\alpha_4\alpha_5\alpha_6\alpha_7} \ 
{i}^{ N_7 N_3 N_8 
N_9\,(a_3)}_{\alpha_7\alpha_3\alpha_8\alpha_9} \ {i}^{ N_9 
N_6 N_2 N_{10}\,(a_4)}_{\alpha_9\alpha_6\alpha_2\alpha_{10}} 
\ {i}^{N_{10} N_8 N_5 
N_1\,(a_5)}_{\alpha_{10}\alpha_8\alpha_5\alpha_1}.
\label{AA}
\end{equation}

\bigskip \noindent{\bf Case B.} Substituting the mode expantion
(\ref{genPHI}) into the definition 
of $\phi$
\begin{eqnarray}\label{eq:canC1}
\phi(g_1\ldots g_4) &=& \int\limits_{SO(4)} dg \ \ 
\tilde\phi(g_1g\ldots g_4g).
\end{eqnarray}
one obtains:
\begin{equation}
\phi(g_1,\ldots,g_4) = \sum_{N_1\ldots N_4} \sum_\Lambda 
\left(\phi^{N_1\ldots N_4}_{\alpha_1\ldots\alpha_4}\ 
D^{(N_1)}_{\alpha_1\beta_1}(g_1) \ldots 
D^{(N_4)}_{\alpha_4\beta_4}(g_4)\ C^{N_1\ldots 
N_4\,\Lambda}_{\beta_1\ldots\beta_4}\right) \ 
\left({C}^{N_1\ldots N_4\,\Lambda}_{\gamma_1\ldots\gamma_4} 
w_{\gamma_1}\ldots w_{\gamma_4}\right).
\end{equation}
The quantity in the second parenthesis is the scalar product 
of two $SO(3)$ invariant vectors in the representation 
$\Lambda$.  Since invariant vectors exist only in simple 
representations, this quantity is non-vanishing only when 
$\Lambda$ is a simple representations $N$.  In this case its 
value is given by (\ref{ultima})
\begin{equation}\label{*}
{C}^{N_1\ldots N_4\,N}_{\gamma_1\ldots\gamma_4} 
w_{\gamma_1}\ldots w_{\gamma_4} = {1\over \sqrt{{\rm 
dim}_{N_1} \ldots {\rm dim}_{N_4} }}.
\end{equation}
This suggests to redefine the mode expansion in term of the 
new fields
\begin{equation}
\phi_{\mathcal{B}}{}^{N_1\ldots 
N_4}_{\alpha_1\ldots\alpha_4} = \phi^{N_1\ldots 
N_4}_{\alpha_1\ldots\alpha_4} \sqrt{{\rm dim}_{N_1} \ldots 
{\rm dim}_{N_4} }
\end{equation}
Substituting this into the mode expansion, we get
\begin{equation}
\phi(g_1,\ldots,g_4) = \sum_{N_1\ldots N_4} 
\phi_{\mathcal{B}}{}^{N_1\ldots 
N_4}_{\alpha_1\ldots\alpha_4}\ 
D^{(N_1)}_{\alpha_1\beta_1}(g_1) \ldots 
D^{(N_4)}_{\alpha_4\beta_4}(g_4)\ S^{N_1\ldots 
N_4}_{\beta_1\ldots\beta_4}.
\end{equation}
Here $S^{N_1\ldots N_4}_{\beta_1\ldots\beta_4}$ is the {\em 
normalized\/} Barrett-Crane intertwiner\footnote{%
Barrett and Crane use a non-normalized expression in 
\cite{Barrett:1998}.  Notice also that they call this 
intertwiner ``vertex''.  Here we reserved the expression 
``vertex'' for the five-valent vertices of the 2-complex 
(the dual to the 4-simplices).  This choice is consistent 
with standard Feynman diagrammatic and with 
\cite{Reisenberger:1997}.}
\begin{equation} \label{bcv}
S^{N_1\ldots N_4}_{\beta_1\ldots\beta_4} := \frac{\sum_N \ 
C^{N_1\ldots N_4 \, N}_{\beta_1\ldots\beta_4}} 
{\sqrt{\sum_{N} \ C^{N_1\ldots N_4\ 
N}_{\beta_1\ldots\beta_4} {C}^{N_1\ldots N_4\ 
N}_{\beta_1\ldots\beta_4}}}\ \ .
\end{equation}
The normalization, given by the denominator in the above 
expression, is in the scalar product of $K_{\vec N_{e}}$.  
Since the quantities $C^{N_1\ldots N_4\ 
N}_{\beta_1\ldots\beta_4}$ are normalized, the denominator 
is the square root of the dimension of the subspace of 
$K_{\vec N_{e}}$ spanned by the intertwiners having an 
intermediate simple representation (See Appendix).

Let us now find the mode expansion of the action.  Using the 
result for the integral of the product of two matrix 
elements, the kinetic term becomes
\begin{equation}
\frac{1}{2} \int \prod_{i=1}^4 dg_i ~ 
~~\phi^2(g_1,\ldots,g_4) = \frac{1}{2} \sum_{N_1\ldots N_4} 
\phi_{\mathcal{B}}{}^{N_1\ldots N_4}_{\alpha_1\ldots\alpha_4 
}\ \phi_{\mathcal{B}}{}^{ N_1\ldots 
N_4}_{\alpha_1\ldots\alpha_4 }.
\end{equation}
The potential term gives
\begin{eqnarray}
&&\frac{\lambda}{5!} \int \prod_{i=1}^{10} dg_i 
~~\phi(g_1,g_2,g_3,g_4)\ \phi(g_4,g_5,g_6,g_7)\ 
\phi(g_7,g_3,g_8,g_9)\ \phi(g_9,g_6,g_2,g_{10}) \ 
\phi(g_{10},g_8,g_5,g_1) \\ \nonumber && = 
\frac{\lambda}{5!} \sum_{N_1\ldots N_{10}} 
\sum_{\Lambda_1\ldots\Lambda_5} \phi_{\mathcal{B}} {}^{ N_1 
N_2 N_3 N_4 }_{\alpha_1\alpha_2\alpha_3\alpha_4} 
\phi_{\mathcal{B}} {}^{ N_4 N_5 N_6 N_7 
}_{\alpha_4\alpha_5\alpha_6\alpha_7} \phi_{\mathcal{B}} {}^{ 
N_7 N_3 N_8 N_9 }_{\alpha_7\alpha_3\alpha_8\alpha_9} 
\phi_{\mathcal{B}} {}^{ N_9 N_6 N_2 
N_{10}}_{\alpha_9\alpha_6\alpha_2\alpha_{10}} 
\phi_{\mathcal{B}} {}^{ N_{10} N_8 N_5 
N_1}_{\alpha_{10}\alpha_8\alpha_5\alpha_1} ~~~~ {\cal 
B}_{N_1,\ldots,N_{10}} ~.
\end{eqnarray}
Here ${\cal B}_{N_1,\ldots,N_{10}}$ is the Barrett-Crane 
vertex-amplitude, which is a $(15-j)$-symbol with 
Barrett-Crane intertwiners
\begin{equation} \label{B2}
{\cal B}_{N_1,\ldots,N_{10}} := {S}^{N_1 N_2 N_3 
N_4}_{\alpha_1\alpha_2\alpha_3\alpha_4}\ {S}^{N_4 N_5 N_6 
N_7}_{\alpha_4\alpha_5\alpha_6\alpha_7}\ {S}^{N_7 N_3 N_8 
N_9}_{\alpha_7\alpha_3\alpha_8\alpha_9}\ {S}^{N_9 N_6 N_2 
N_{10}}_{\alpha_9\alpha_6\alpha_2\alpha_{10}}\ {S}^{N_{10} 
N_8 N_5 N_1}_{\alpha_{10}\alpha_8\alpha_5\alpha_1}.
\label{BB}
\end{equation}

%%%%%%%%%%%%%%%%%%%%%%%%%%%%%%%%%%%%%%%%%%%%%%%%%%%%%%%%%
%% %% S E C T I O N %%
%%%%%%%%%%%%%%%%%%%%%%%%%%%%%%%%%%%%%%%%%%%%%%%%%%%%%%%%

\section{Feynman graphs}
\label{sec:feyn}

%%%%%%%%%%%%%%%% CASE A %%%%%%%%%%%%%%%%%%%%%%%%

\noindent{\bf Case A.} The partition function is given by 
the integral over modes
\begin{equation}
Z = \int \big[ D\phi_{{\mathcal {A}}}^{N_1\ldots 
N_4\,\Lambda}\big] \ \ e^{-S[\phi_{\mathcal {A}}]}.
\end{equation}
We expand $Z$ in powers of $\lambda$.  The Gaussian 
integrals are easily computed, giving the propagator 
\begin{equation}\label{prop-a}
P_{\mathcal{A}}^{N_{1}\ldots N_{4}\,\Lambda,N'_{1}\ldots 
N'_{4}\,\Lambda'}\ =: \ <\phi_{\mathcal {A}}^{N_1\ldots 
N_4\,\Lambda},\ \phi_{\mathcal {A}}^{N'_1\ldots 
N'_4\,\Lambda'}>\ = \ \frac{1}{4!} \sum_{\sigma}\ \ \delta^{ 
N_1 N'_{\sigma(1)}} \ldots \delta^{ N_4 N'_{\sigma(4)}}\ 
M_{\sigma}^{N_{1}\ldots N_{4}\,}{}^{\Lambda}_{\Lambda'}
\end{equation}
$\sigma$ are the permutations of $\{1,2, 3,4\}$.  The matrix 
$M_{\sigma}$, defined in the appendix, gives the change of 
basis in $K_{\vec N_{e}}$ that corresponds to a permutation 
of the four representations (see appendix).  There is a 
single vertex, of order five, which is:
\begin{equation}\label{vert-a}
<\phi_{\mathcal {A}}^{N_1 N_2 N_3 N_4\,\Lambda_1}\ \ldots \ 
\phi_{\mathcal {A}}^{N_{10} N_8 N_5 N_1\,\Lambda_5}>\ = \ 
\lambda \ \ {\mathcal 
{A}}_{N_1,\ldots,N_{10},\Lambda_1,\ldots,\Lambda_5}.
\end{equation}

%%%%%%%%%%%%%%%% CASE B %%%%%%%%%%%%%%%%%%%%%%%%

\bigskip \noindent{\bf Case B.} The partition function in 
this case is given by
\begin{equation}\label{def:Z}
Z = \int \big[ D\phi_{\mathcal{B}}{}_{\alpha_1\ldots 
\alpha_4}^{N_1\ldots N_4}\big]\ \ 
e^{-S[\phi_{\mathcal{B}}{}]}.
\end{equation}
The propagator is
\begin{equation}\label{prop-b}
P_{\mathcal{B}}{}^{N_{1}\ldots N_{4},\ N'_{1}\ldots N'_{4} 
}_{\alpha_{1}\ldots \alpha_{4},\ \ \alpha'_{1}\ldots 
\alpha'_{4}}=: <
\phi_{\mathcal {B}}{}^{N_1\ldots N_4}_{\alpha_1\ldots \alpha_4}, 
\phi_{\mathcal {B}}{}^{N'_1\ldots N'_4}_{\alpha'_1\ldots \alpha'_4}
>\ = \ \frac{1}{4!} \sum_{\sigma} \delta^{ 
N_1 N'_{\sigma(1)}} \ldots \delta^{ N_4 N'_{\sigma(4)}}\ \ 
\delta^{\alpha_1\alpha'_{\sigma(1)}}\ldots 
\delta^{\alpha_4\alpha'_{\sigma(4)}}
\end{equation}
The vertex is given by:
\begin{equation}
<\phi_{\mathcal{B}}{}_{\alpha_1\alpha_2\alpha_3\alpha_4}^{ 
N_1 N_2 N_3 N_4} \ldots 
\phi_{\mathcal{B}}{}_{\alpha'_{10}\alpha'_8\alpha'_5\alpha'_1}^{ 
N_{10} N_8 N_5 N_1}>\ \ = \ \ \lambda \ \ 
\delta^{\alpha_1\alpha'_1}\ldots 
\delta^{\alpha_{10}\alpha'_{10} } \ \ {\mathcal 
{B}}_{N_1\ldots N_{10}} ~~,
\end{equation}

\bigskip

\begin{figure}
\centerline{ \hbox{\psfig{figure=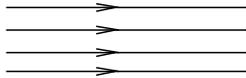,height=1cm}} 
} \bigskip \caption{The propagator can be represented by a 
collection of four strands, each carrying a simple 
representation.}
\label{fig:prop}
\end{figure}

The set of Feynman rules one gets is similar in the two 
cases.  First, we obtain the usual overal factor 
${\lambda^{v[\Gamma]}/{\rm sym}[\Gamma]}$ (see for instance 
\cite{Peskin}, page 93).  Next, we represent each of the 
terms in the right hand side of the definitions 
(\ref{prop-a},\ref{prop-b}) of the propagator by four 
parallel strands, as in Fig.\ \ref{fig:prop}, carrying the 
indices at their end.  Then, we can represent the propagator 
itself by the symmetrization of the four strands.  In 
addition, in case A, edges $e$ are labeled by a 
representation $\Lambda_e$ (not necessarily simple).

\begin{figure}
\centerline{ \hbox{\psfig{figure=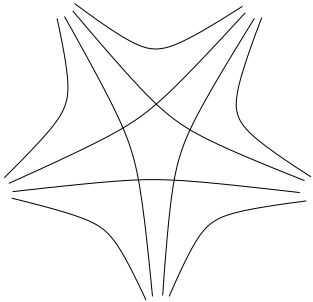,height=4cm}}} 
\bigskip \caption{The structure of the vertex of the fat 
5-valent graphs generated by the Feynman expansion.}
\label{fig:vertex}
\end{figure}

The Feynman graphs we gets are all possible ``4-strand'' 
5-valent graphs, where a ``4-strand graph'' is a graph whose 
edges are collections of four strands, and whose vertices 
are the ones shown in Fig.  \ref{fig:vertex}.  Each strand 
of the propagator can be connected to a single strand in 
each of the five ``openings'' of the vertex.  Orientations 
in the vertex and in the propagators should match (this can 
always be achieved by changing a representation to its 
conjugate).  Each strand of the 4-strand graph, goes through 
several vertices and several propagators, and then closes to 
itself, forming a cycle.  Note that a particular strand can 
go through a particular edge of the 4-strand graph more than 
once.  Cycles get labeled by the simple representations of 
the indices.  For each graph, the abstract set formed by the 
vertices, the edges, and the cycles forms a 2-complex, in 
which the faces are the cycles.  The labeling of the cycles 
by simple representations of $SO(4)$ determines a coloring 
of the faces by spins.  Thus, we obtain a colored 2-simplex, 
namely a spin foam.

In Case A, edges $e$ are labeled by an intertwiner with 
index $\Lambda_e$ (in the basis (\ref{basis1})).  Vertices 
$v$ contribute a factor $\lambda$ times ${\mathcal {A}}$, 
(see (\ref{A})), which depends on ten simple representations 
labeling the cycles that go through the vertex, and on five 
intertwiners, basis elements in $K_{\vec N_{e}}$, labeling 
the edges that meet at $v$.  Each edge contracts two 
vertices, say $v$ and $v'$.  It contributes a matrix 
$M_{\sigma}$.  This is the matrix of the change of basis 
from the basis in $K_{\vec N_{e}}$ used at $v$ and the basis 
used at $v'$.  If we fix a basis of intertwiners $i_{e}$ in 
$K_{\vec N_{e}}$ for every $e$, once and for all for each 
fixed 2-complex --that is, for each assigned permutation in 
(\ref{prop-b})-- and use vertex amplitude $A(\vec N_{v}, 
\vec i_{v})$ given in (\ref{AA}), the matrix $M_{\sigma}$ is 
absorbed into the vertex amplitude and the propagator is the 
identity.  The weight of 2-complex $\Gamma$ is then given by 
(\ref{case-a}).

In Case B, there are no labeling of edges, and we gets an 
additional contribution from the summing of the $\alpha's$ 
indices of the Kroneker deltas $\delta^{\alpha\alpha'}$ 
around each cycle.  This gives a factor ${\rm dim}_{N_f}$ 
for every face $f$.  Thus, faces $f$ are labeled by simple 
representations $N_f$ and contribute a factor ${\rm 
dim}_{N_f}$.  Vertices contribute a factor $\lambda$ times 
${\mathcal {B}}$, where ${\mathcal {B}}$ is the 
Barrett-Crane weight (\ref{B}), which depends on the ten 
simple representations of the faces adjacent to the vertex.  
This yields the result (\ref{case-b}).

\section{Triangulations, orientation and symmetries of 
$\phi$}
\label{sec:triang}

In this section we comment on the relation between 
2-complexes and triangulations, and we discuss some possible 
modification of our models, which leads to oriented 
2-complexes.
\begin{itemize}
\item
Given a triangulation $\Delta$, the (abstract combinatorial 
structure of its) dual 2-skeleton, $\Gamma(\Delta)$ is a 
2-complex.  In general, not every 2-complex comes from a 
triangulation triangulation of a 4-manifold in this way.  In 
3d, one can nicely characterize the 2-complexes that are 
derived from triangulations \cite{Petronio}.  Since this 
characterization introduces an interesting technology, we 
mention it here, adapted from \cite{Petronio}.

As we have seen, a 2-complex corresponds to a 4-strand 
graph.  Consider a sequence of edges forming a closed loop 
in the graph, starting with an edge $e$.  Label the strands 
of $e$, say with integers (1,2,3 in 3d, and 1,2,3,4 in 4d).  
The labeling can be carried over to the next edge in the 
loop, say $e'$ by noticing that the strands of $e'$ are 
naturally paired with the strands of $e$: a strand in $e$ 
and a strand in $e'$ are paired either if they are the same 
strand that continues across the vertex, or if they both 
continue across the vertex into an edge $e''$.  Thus, the 
labeling can be carried over the loop.  In closing the loop, 
we obtain a new labeling of the strands of $e$.  If the new 
labeling is an even (respectively odd) permutation of the 
original labeling, we say that the loop is even 
(respectively odd).

In 3d we have the following result \cite{Petronio}.  If, and 
only if, each cycle (a loop that bounds a face) of the 
2-complex is even, we can ``thicken'' faces, edges and 
vertices of the 2-complex, obtaining a 3 manifold with 
boundaries, whose spine is the 2-complex .  If, in addition, 
each component of the boundary of this manifold is a sphere, 
we can then fill the sphere with a ball, and obtain a 
triangulated manifold, whose dual 2-skeleton is the 
2-complex.  We are not aware of related results in 4d.

\item
Next, let us now consider the notion of orientation.  The 
notion of orientability of a manifold can be extended to 
abstract, $\,n$-dimensional triangulations.  In $n$ 
dimensions, an $n\,$-simplex $S$ has $n+1$ points 
($0$-simplices) in its boundary.  An orientation of $S$ is a 
choice of an ordering of these $n+1$ points, up to even 
permutations.  An $n\,$-simplex is bounded by n+1 
($n\,$-1)-simplices.  Each of these is bounded by n-2 
points, obtained by dropping one of the points of the 
$n$-simplex.  Each ($n\,$-1)-simplex inherits an 
(``outgoing'') orientation from the $n$-simplex.  This can 
be obtained by dropping the missing point in an even 
ordering of the points of the ($n\,$-1)-simplices in which 
this missing point is first.  Two $n\,$-simplices that share 
an ($n\,$-1)-simplex have consistent orientation if the 
shared ($n\,$-1)-simplex inherits opposite orientations from 
the two $n\,$-simplices it bounds.  A triangulation is 
orientable if it admits a consistent orientations of all its 
$n$-simplices.  Clearly, the triangulation of an orientable 
manifold is orientable.

Now, the notion of orientability extends to 2-complexes with 
4-valent edges and 5-valent vertices.  In fact, notice that 
in the boundary of an $n\,$-simplex of a triangulation, the 
($n\,$-1)-simplices are paired with the points (the point 
that does not belong to the ($n\,$-1)-simplex).  Thus, the 
ordering of the points corresponds to an ordering of the 
($n\,$-1)-simplices.  Consequently, we can simply define the 
orientation of a vertex of a 2-complex as an ordering of its 
adjacent edges, up to even permutations.  In turn, the 
orientation of a vertex induces an (outgoing) orientation of 
its adjacent edges, namely an ordering of the faces, 
obtained as above.  The orientation of two vertices 
separated by an edge is consistent if the edge inherits 
opposite orientations from the two vertices.  The 2-complex 
is orientable if it admits a consistent orientation of all 
its vertices.  Clearly, a 2-complex is orientable if all its 
loops (not just the cycles) are even.  A 2-complex derived 
from a triangulation of an orientable manifold, is 
orientable.

It is not difficult to modify our field theory in such a way 
that it generates only orientable 2-complexes.  To do that, 
it is sufficient to replace the requirement that $\phi$ is 
symmetric under any permutation of its arguments, eq.  
(\ref{perm}) with
\begin{equation}
\label{permE}
\phi(g_{1},g_{2},g_{3},g_{4})\ =\ 
\phi(g_{\sigma_{E}(1)},g_{\sigma_{E}(2)}, 
g_{\sigma_{E}(3)},g_{\sigma_{E}(4)}),
\end{equation}
where $\sigma_{E}$ is an {\em even\/} permutation, and to 
rewrite the action as
\begin{eqnarray}
S[\phi] &=& \frac{1}{2} \int \prod_{i=1}^{4} dg_i ~ 
\phi(g_1,g_2,g_3,g_4)\ \phi(g_3,g_2,g_1,g_4) \ + 
\frac{\lambda}{5!} \int \prod_{i=1}^{10} dg_i 
~~\phi(g_1,g_2,g_3,g_4)\ 
\\ \nonumber
&& \phi(g_4,g_5,g_6,g_7)\ \phi(g_7,g_3,g_8,g_9)\ 
\phi(g_9,g_6,g_2,g_{10}) \ \phi(g_{10},g_8,g_5,g_1).
\end{eqnarray}
(This action is the same as (\ref{action}) if $\phi$ is 
symmetric under any permutation of its arguments.)\ If we do 
so, the propagators (\ref{prop-a}) and (\ref{prop-b}) 
contain odd permutations only.  A moment of scrutiny of the 
vertex shows that the pairing of the strands discussed above 
(in defining the parity of the loops) determines always an 
odd permutation.  If we consider a loop of edges in the 
Feynman diagram, we cross an equal number of vertices and 
edges.  Therefore the strands undergo an even number of odd 
permutations.  Therefore in closing the loop we obtain an 
even permutation.  Therefore all loops are even and the 
2-complex is orientable.

\item
Other versions of the model can be defined by considering 
more complicated symmetry properties of the function $\phi$.  
In particular, there are 5 irreducible representation of the 
permutation group of 4 elements $\Sigma_4$.  Two of these 
are one-dimensional (the trivial and the signature), and the 
other three have dimensions 2,3 and 5.  If we denote 
$\chi_i(g), i =1,\cdots 5 $ the character of these 
representations, we can define five different fields
\begin{equation}
        \phi_i(g_1,g_2,g_3,g_4) = {1\over 4!} \sum_{\sigma \in 
        \Sigma_4} \chi_i(\sigma) 
        \phi_i(g_{\sigma(1)},g_{\sigma(2)},g_{\sigma(3)},g_{\sigma(4)}).
\end{equation}
Using the orthonormality of the characters, the original 
field is totally determined by the $\phi_i$
\begin{equation}
        \phi =\sum_i \chi_i(1) \phi_i.
\end{equation}
Taking the field $\phi$ to be one of the $\phi_i$ leads to 
five different models having slightly different Feynman 
rules.  The propagator associated with permutation gets 
multiplied by the character of this permutation in the given 
representation, and so on.  Finally, one can perhaps add the 
actions of these five models, and use the coupling constants 
to try to control the topologies appearing in the Feynman 
expansion.
\end{itemize}

\section{Conclusion}
\label{sec:disc}

We close with some general comments.
\begin{itemize}
\item The results of this paper generalize easily to higher 
dimensions.  In dimension $D$, we replace the group ${\rm 
SO}(4)$ with ${\rm SO}(D)$, the sphere $S^3$ with 
$S^{D-1}={\rm SO}(D)/{\rm SO}(D-1)$.  The field $\phi$ is 
defined over $D$ copies of the $SO(D)$.  The interaction 
term is of order $D+1$ and has the structure of a 
$D$-simplex.  The simple representations of ${\rm SO}(D)$ 
are discussed in details in \cite{Freidel:1999b}.  In any 
dimension, Case B yields the higher-dimensional 
generalizations of Barrett-Crane model discussed in 
\cite{Freidel:1999b}.
\item
The sum over spin foam we have obtained is presumably 
divergent.  There are two sources of divergences.

The first is the sum over the colorings.  This divergence 
can be regulated by replacing the Clebsch-Gordan 
coefficients of the group with the ones of the corresponding 
quantum group, with $q^k=1$, as in the Barrett-Crane model.  
This procedure cuts off the sum over representations, and 
thus makes this sum finite.  In this regard, we think that 
it would be interesting to explore the construction of the 
field theory presented here, but over a non-commutative 
3-sphere.

The second source of divergences is the sum over 
2-complexes.  The number $N$ of diff-inequivalent 
triangulations of a given four dimensional manifold grows 
exponentially only with the number $n$ of simplices 
\cite{Carfora}.  That is $N(n) < exp \{\alpha n\}$ for large 
$n$.  This fact suggests that the sum might converge for 
$\lambda$ smaller than a critical value, determined by 
$\alpha$ \cite{Roberto}.  On the other hand, the Feynman 
expansion generates triangulations of arbitrary manifolds, 
and this might destroy the convergence.
\item
The Barrett-Crane model has been obtained by ``quantizing a 
geometric 4-simplex'' \cite{Barrett:1998}.  That is, by 
realizing certain geometrical quantities, as the 
``bivectors'' that can be associated to the faces of the 
tetrahedron, as operators.  Simple representations emerge in 
this process: in the quantum theory, the condition on 
bivectors that they describe planes in $R^4$ translates into 
the the simplicity condition (\ref{simplicity}) on the 
representation.  In turn, the condition that these planes 
intersect along lines determines the Barrett-Crane 
intertwiner (see also \cite{Mike}).  The model described 
here as Case A can be seen as a different solution to this 
intersection constraint.  In detail, consider a single 
tetrahedron of a triangulation.  The tetrahedron belongs to 
two different 4-simplices.  In the Barrett-Crane model, one 
``quantizes each 4-simplex'' separately.  The corresponding 
intersection constraint is imposed two times.  If, 
alternatively, we view the tetrahedron as a single element 
of the triangulation, we may expect a single constraint 
corresponding to the requirement that the two faces 
intersect on a line.  The vertex of the model A satisfies 
the intersection constraint in this sense.  Indeed, it 
satisfies the identity pictured in Fig.  \ref{fig:id}.
\begin{figure}
\centerline{\hbox{\psfig{figure=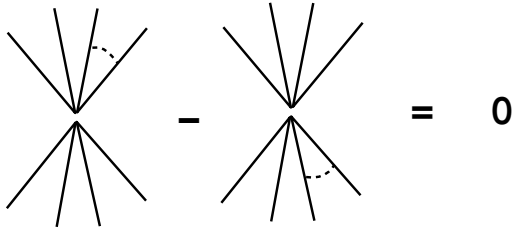,height=3cm}}}
\bigskip \caption{The identity satisfied by the 4-valent 
vertices of model A. The sum over all representations 
$\Lambda_e$ is implicit.}
\label{fig:id}
\end{figure}
Thus, the model of case A provides another solution to the 
intersection constraints.

\item
We close with a comment on the physical interpretation of 
the theory we have described.  Consider the model of case A. 
The field is $\phi^{N_1\ldots N_4\,\Lambda}$.  The indices 
corresponds to the assignment of a (simple) representations 
to each face of a tetrahedron, and of an (arbitrary) 
intertwiner to the bulk of the tetrahedron.  This assignment 
can be thought as the assignment of metric properties to the 
tetrahedron: indeed, we know from loop quantum gravity 
\cite{loop,VolLoop} and from other approaches 
\cite{Barbieri}, that the $N_i$'s are naturally related to 
the area of the faces, and their intertwiner to the volume.  
[More precisely, the intrinsic geometry of a classical 
tetrahedron in 4d is determined by six numbers (4x4 
coordinates of the vertices, minus ten dimensions from the 
Poincar\'{e} group), but two of these quantities do not 
commute as quantum operators and therefore only five quantum 
numbers fix the state.]  Therefore, we can view 
$\phi^{N_1\ldots N_4 \,\Lambda}$ as a quantum amplitude for 
a certain geometry of a tetrahedron or, more generally, of 
an elementary ``chunk'' of space, with a certain volume, and 
bounded by four elementary surfaces with a certain area.  
Correspondingly, we can view the field $\phi(x_1\ldots x_4)$ 
as the ``Fourier transform'' of such amplitude.  From this 
perspective, the quantum theory defined by the action 
(\ref{action}) is the second quantization of the quantum 
theory of the geometry of an elementary chunk of space.  As 
a second quantized theory, it is a ``multiparticle theory'', 
in which particles are created and destroyed.  Four 
dimensional spacetime can thus be viewed as a Feynman 
history of creations, destruction and interactions of these 
elementary quanta of space.
\end{itemize}

\vskip1cm
\centerline{--------------------------------}
\vskip1cm

We thank Mauro Carfora, Louis Crane, Carlo Petronio and 
Annalisa Marzuoli for discussions and correspondence.  
R.D.P. is supported by a Dalla Riccia Fellowship.  K.K. was 
supported in part by a Braddock fellowship from Penn State.  
This work was partially supported by NSF Grants PHY-9900791, 
PHY 95-15506, PHY94-07194, PHY95-14240 and by the Eberly 
research founds of Penn State University.

\appendix

\section{Some facts on representation theory}

We collect here some facts on the representation theory of 
$SO(D)$.  We label finite dimensional irreducible 
representation of $SO(D)$ by their highest weight $\Lambda$.  
$\Lambda$ is a vector of length $n=[D/2]$ ($[\cdot ]$ is the 
integral part): $\Lambda =( N_1,\cdots, N_n)$, where $N_i$ 
are integer and $N_1\geq \cdots \geq N_n $.  If we are 
interested in representations of $Spin(D)$ we let the $N_i$ 
be half integers.  The representation labeled by the highest 
weight $\Lambda=(N, 0,\cdots,0)$ are called {\it simple} or 
{\it spherical}.  Let $X_{ij}, 1\leq i,j \leq D$ be a basis 
of the Lie algebra of $SO(D)$.  The simple representation 
are the ones for which the ``simplicity'' relations
\begin{equation}
        X_{[ij}X_{ij]} \cdot V_N =0
        \label{simplicity}
\end{equation}
are satisfied.  Here, $[\cdot ] $ denotes the 
antisymmetrisation.  The representation space $V_N$ of a 
simple representation can be realized as a space of 
spherical harmonics, that is, harmonic homogeneous 
polynomial on $R^D$.  Any $L^2$ function on the sphere can 
be uniquely decomposed in terms of these spherical harmonics
\begin{equation}
        {\cal L}^2( S^{D-1}) = \oplus_{N=0}^{\infty} V_N.
\end{equation}

In the case of $SO(4)$, since $Spin(4) = SU(2) \times 
SU(2)$, there is an alternate description of the 
representation as products of two representations $j'$ and 
$j''$ of $SU(2)$.  The relation with the highest weight 
presentation is given by
\begin{equation}
        N_{1}= j'+j'', \ \ \ \ \ N_{2}= j'+j''.
\end{equation}
the simple representation are therefore the representation 
in which $ j'=j'':= j$.  Thus, we can label simple 
representations with a half integer spin $j$.  Notice that 
the integer ``color'' $N=2j$ is also the (nonvanishing 
component of the) highest weight of the representation.

An elementary illustration of these simple representations 
can be given as follows.  The vectors $v^\alpha$ of the 
representation $\Lambda=(j',j'')$ can be written as spinors 
$\psi^{A_1\ldots A_{j'}\, \dot B_1\ldots \dot B_{j''}}$ with 
$j'$ ``undotted'' symmetrized indices $A_i=1,2$, 
transforming under one of the $SU(2)$, and $j''$ ``dotted'' 
antisymmetrized indices $\dot B_i=1,2$ transforming under 
the other.  Consider the particular vector $w^\alpha$ which 
(in the spinor notation, and in a given basis) has 
components $\psi^{A_1\ldots A_{j'}\, \dot B_1\ldots \dot 
B_{j''}}=\epsilon^{(A_1\dot B_1}\ldots \epsilon^{A_j) \dot 
B_j}$, where $\epsilon^{A\dot B}$ is the unit antisymmetric 
tensor, and the symmetrization is over the $A_i$ indices 
only.  The subgroup of ${\rm SO}(4)$ that leaves this vector 
invariant is a ${\rm SO}(3)$ subgroup of ${\rm SO}(4)$ 
(which depends on the basis chosen).  Clearly, since 
$\epsilon^{A\dot B}$ is the only object invariant under this 
${\rm SU}(2)$, a normalized ${\rm SO}(3)$ invariant vector 
(and only one) exists in these simple representations only.

Equivalently, the simple representations of ${\rm SO}(4)$ 
are the ones defined by the completely symmetric traceless 
4d tensors of rank $N$.  The invariant vector $w$ is then 
the traceless part of the tensor with (in the chosen basis) 
all component vanishing except $w^{444...}$.  The ${\rm 
SO}(3)$ subgroup is given by the rotations around the fourth 
coordinate axis.  The relation between the vector and spinor 
representation is obtained contracting the spinor indices 
with the (four) Pauli matrices: 
$v^{\mu_1\ldots\mu_j}=\psi^{A_1\ldots A_{j'}\, \dot 
B_1\ldots \dot B_{j''}}\sigma^{\mu_1}_{A_1\dot B_1}\ldots 
\sigma^{\mu_j}_{A_j\dot B_j}$.

In the main text we have used the following properties of 
the simple representations
\begin{itemize}
\item Let $V_\Lambda$ be a representation of $SO(D)$, we say 
that $\omega \in V_\Lambda$ is a spherical vector if it is 
invariant under the action of $SO(D-1)$.  Such a vector 
exists if and only if the representation is simple.  In that 
case this vector is unique up to normalization.

Let $\omega$ be a vector of $V_\Lambda$, and consider an 
orthonormal basis $v_i$ of $V_\Lambda$.  We can construct 
the following functions on $G$:
\begin{equation}
        \Theta_i(g) = <\omega | D^{-1}(g) | v_i>.
\end{equation}
These functions span a subspace of ${\cal L}^2(G)$.  The 
group acts on this subspace by the right regular 
representation, and the corresponding representation is 
equivalent to the representation $V_\Lambda$.  If $\omega$ 
is spherical, then these functions are in fact ${\cal L}^2$ 
functions on the quotient space $ SO(D)/SO(D-1) =S^{D-1}$ 
and $V_\Lambda$ is therefore a spherical representation.  On 
the other hand, if the representation is spherical then we 
can construct a spherical vector: $\Theta_{\omega}(g) = 
\sum_i \Theta_i(g) \Theta_i(1)$.  When $\omega $ is 
spherical, the spherical function $\Theta_{\omega}$ is a 
function on the double quotient space $SO(D-1)\backslash{\rm 
SO}(D)/SO(D-1) = U(1)$.  It is now a standard exercise to 
show that there is a unique harmonic polynomial on $R^D$ 
invariant by $SO(D-1)$ of a given degree, hence there is a 
unique spherical function.
\item
The space of intertwiner of three representations of $SO(4)$ 
is at most one dimensional.

The dimension $n_{\Lambda_1, \Lambda_2, \Lambda_3}$ of the 
space of intertwiner between three representations 
$\Lambda_1, \Lambda_2, \Lambda_3$ is given by the integral
\begin{equation}
        n_{\Lambda_1, \Lambda_2, \Lambda_3} =\int dg\ 
        \chi_{\Lambda_1}(g) 
        \chi_{\Lambda_2}(g)\chi_{\Lambda_3}(g)
\end{equation}
where $\chi_\Lambda$ are the characters of the 
representation $\Lambda$.  For $SO(4)$, since any 
representation is the product of two representation of 
$SU(2)$ the intertwining number of three $SO(4)$ 
representation is the product of $SU(2)$ intertwining 
numbers.  It is well known that this number is 0 or 1 for 
$SU(2)$.

\item
The representation of $SO(N)$ are real.  This means that it 
is always possible to choose a basis of $V_\Lambda$ such 
that the representation matrices are real.  If we are 
interested by representation of half integer spin of 
$Spin(N)$ it is still true that $\Lambda $ is equivalent to 
its complex conjugate or dual.  However the isomorphism is 
non trivial.  The difference between this two case can be 
characterized by the value of the following integral
\begin{equation} \label{test}
I_\Lambda = \int dg \ \chi_\Lambda(g^2).
\end{equation}
This is $+1$ if $\Lambda$ is an integer spin representation 
and $-1$ if $\Lambda$ is a half-integer spin representation.  
This is easily seen in the case of $Spin(4)$, since any 
representation is the product of two representations of 
$SU(2)$ the integral (\ref{test}) reduces to the product of 
two $SU(2)$ integrals.
\end{itemize}

\section{Intertwiners and their spaces}

The integral of the product of two unitary matrix elements 
is given by:
\begin{equation}
\int_{SO(4)} dg\ \overline{D^{(\Lambda)}_{\alpha\beta}(g)} 
D^{(\Lambda')}_{\alpha'\beta'}(g) = {1\over {\rm 
dim}_\Lambda} \delta^{\Lambda\Lambda'}\ 
\delta_{\alpha\alpha'}\ \delta_{\beta\beta'}.
\label{app:2-int}
\end{equation}
Since we choose a real basis, the bar can be dropped.  The 
integral of the product of three group elements is:
\begin{equation} \label{app:3-int}
 \int\limits_{SO(4)} dg \ D^{(N_1)}_{\alpha_1\beta_1}(g) 
 D^{(N_2)}_{\alpha_2\beta_2}(g) 
 {D^{(\Lambda)}_{\alpha\beta}(g)} = C^{N_1 
 N_2\,\Lambda}_{\alpha_1\alpha_2\alpha}\ \ {C}^{N_1 
 N_2\,\Lambda}_{\beta_1\beta_2\beta}.
\end{equation}
Here $C^{N_1 N_2 \Lambda}_{\gamma_1\gamma_2\gamma}$ are 
normalized $(3-j)$-symbols for ${\rm SO}(4)$.  Several 
normalization conditions are used in the literature.  We 
follow here the most common one, in which the value of the 
$\theta$-symbol is one.  That is $C^{N_1 
N_2\,\Lambda}_{\alpha_1\alpha_2\alpha}\ \ {C}^{N_1 
N_2\,\Lambda}_{\alpha_1\alpha_2\alpha}=1$.  The intertwiner 
from the tensor product of two representations $N_1,N_2$ to 
a representation $\Lambda$, if it exists is unique.

Next, consider {\em four\/} representations $N_{1}\ldots 
N_{4}$.  They are defined on the Hilbert spaces $H_{1}\ldots 
H_{4}$.  Consider the tensor product $H_{N_{1}\ldots 
N_{4}}=H_{1}\otimes \ldots \otimes H_{4}$.  This space 
decomposes into irreducibles.  In particular, it contains 
the trivial representation, with a certain multiplicity $m$.  
We denote the $m$ dimensional subspace of $H_{N_{1}\ldots 
N_{4}}$ formed by the trivial representations, that is, the 
$SO(4)$ invariant subspace of $H_{N_{1}\ldots N_{4}}$ as 
$K_{N_{1}\ldots N_{4}}$.  The space $K_{N_{1}\ldots N_{4}}$, 
its scalar product, and its subspaces play a key role in the 
spin foam models.  When the representations $N_{1}\ldots 
N_{4}$ are associated to the four edges adjacent to the edge 
$e$, we write $K_{N_{1}\ldots N_{4}}$ also as $K_{\vec 
N_{e}}$.  The vectors in $K_{N_{1}\ldots N_{4}}$ are the 
``intertwiners'' between the representations $N_{1}\ldots 
N_{4}$.  They are $SO(4)$ invariant tensors with four 
indices, one in each representation $H_{i}$.  We write them 
as $V^{N_1\ldots N_4}_{\gamma_1\ldots \gamma_4}$.  An 
orthonormal basis in $K_{N_{1}\ldots N_{4}}$ can be obtained 
as follows.  We pair the representations as 
$(N_1,N_2),(N_3,N_4)$.  Then we define
\begin{equation}
C^{N_1\ldots N_4\,\Lambda}_{\gamma_1\ldots \gamma_4}= \ 
\sqrt{{\rm dim}_\Lambda} \ C^{N_1 N_2 
\Lambda}_{\gamma_1\gamma_2\gamma}\ {C}^{ N_3 N_4 
\Lambda}_{\gamma_3 \gamma_4 \gamma}.
\label{basis1}
\end{equation}
As $\Lambda$ runs over the finite number of representations 
for which the $(3-j)$-symbols do not vanish, the 
$C^{N_1\ldots N_4\,\Lambda}_{\gamma_1\ldots \gamma_4}$ form 
an orthonormal basis of $K_{N_{1}\ldots N_{4}}$.  The factor 
$\sqrt{{\rm dim}_\Lambda}$ normalizes these vectors in 
$K_{N_{1}\ldots N_{4}}$.  Clearly, there are other bases of 
this kind, obtained by pairing the indices in a different 
manner.  For example, we can we pair the indices as 
$(N_1,N_2),(N_3, N_4)$ and define the basis
\begin{equation}
\tilde C^{N_1 N_{2} N_{3} 
N_4\,\Lambda}_{\gamma_1\gamma_{2}\gamma_{3} \gamma_4}\ = \ 
C^{N_1 N_{3} N_{2} 
N_4\,\Lambda}_{\gamma_1\gamma_{3}\gamma_{2} \gamma_4}.
\end{equation}
Since both the $C$'s and the $\tilde C$'s are orthonormal 
bases, the transformation matrix $M$ is immediately given by 
linear algebra
\begin{eqnarray}
\tilde C^{N_1 N_{2} N_{3} N_4\,\Lambda}_{\gamma_1 \gamma_{2} 
\gamma_{3} \gamma_4}\ \ &=& \sum_{\Lambda'} M^{N_1 N_{2} 
N_{3} N_4\,}{}^{\Lambda}_{\Lambda'} \ C^{N_1 N_{2} N_{3} 
N_4\,\Lambda'}_{\gamma_1 \gamma_{2} \gamma_{3} \gamma_4} \\ 
\nonumber M^{N_1 N_{2} N_{3} N_4\,}{}^{\Lambda}_{\Lambda'} 
&=& C^{N_1 N_{2} N_{3} N_4\,\Lambda}_{\gamma_1 \gamma_{2} 
\gamma_{3} \gamma_4}\ \tilde C^{N_1 N_{2} N_{3} 
N_4\,\Lambda'}_{\gamma_1 \gamma_{2} \gamma_{3} \gamma_4} \\
&=& C^{N_1 N_{2} N_{3} N_4\,\Lambda}_{\gamma_1 \gamma_{2} 
\gamma_{3} \gamma_4}\ C^{N_1 N_{3} N_{2} 
N_4\,\Lambda'}_{\gamma_1 \gamma_{3} \gamma_{2} \gamma_4}.
\label{M}
\end{eqnarray}
In fact, $M^{N_1 N_{2} N_{3} N_4}{}^{\Lambda}_{\Lambda'}$ is 
a $6-j$ symbol for $SO(4)$.  For a generic permutation 
$\sigma$ of four elements, we have a basis
\begin{equation}
{}^\sigma\!  C^{N_1 N_{2} N_{3} 
N_4\,\Lambda}_{\gamma_1\gamma_{2}\gamma_{3} \gamma_4}\ = \ 
C^{N_{\sigma(1)} N_{\sigma(2)} N_{\sigma(3)} 
N_{\sigma(4)}\,\Lambda}_{\gamma_{\sigma(1)}\gamma_{\sigma(2)}\gamma 
_{\sigma(3)} \gamma_{\sigma(4)}}
\end{equation}
and a corresponding matrix of change of basis 
$M_{\sigma}^{N_1 N_{2} N_{3} N_4\,}{}^{\Lambda}_{\Lambda'}$.

Using this technology, the integral of the product of four 
group elements is simply a resolution of the identity in 
$K_{N_{1}\ldots N_{4}}$ and can be written (for any choice 
of basis) as
\begin{equation} \label{app:4-int}
 \int\limits_{SO(4)} dg \ D^{(N_1)}_{\alpha_1\beta_1}(g) 
 \ldots D^{(N_4)}_{\alpha_4\beta_4}(g) = \sum_\Lambda \ \ 
 C^{N_1\ldots N_4\,\Lambda}_{\alpha_1\ldots \alpha_4}\ \ 
 {C}^{N_1\ldots N_4\,\Lambda}_{\beta_1\ldots \beta_4}.
\end{equation}

Finally, we derive equation (\ref{*}) used in the main text.  
Using the ambiguity in the definition of the $(3-j)$-symbol, 
which so far is defined up to a phase factor, we can choose 
it so that the quantity $$C^{N_1 N_2 
N}_{\gamma_1\gamma_2\gamma} w_{\gamma_1} w_{\gamma_2} 
w_{\gamma}$$
is real and positive.  Then its square is given by the 
following integral \begin{equation}\nonumber \left( C^{N_1 
N_2 N}_{\gamma_1\gamma_2\gamma} w_{\gamma_1} w_{\gamma_2} 
w_{\gamma} \right)^2 = \int dg\, D^{(N_1)}_{00} 
D^{(N_2)}_{00} \overline{D^{(N)}_{00}}.
\end{equation}
This integral is computed for the general case of ${\rm 
SO}(N)$ in \cite{Vilenkin:1993b}, Chapter 9.4.11.  In the 
case of ${\rm SO}(4)$ it is given by: 
\begin{equation}\nonumber {1\over (N_1+1)(N_2+1)(N+1)}.
\end{equation}
It follows
\begin{equation}
{C}^{N_1\ldots N_4\,N}_{\gamma_1\ldots\gamma_4} 
w_{\gamma_1}\ldots w_{\gamma_4} = \sqrt{{\rm dim}_N} \ 
\left( {C}^{N_1 N_2 N}_{\gamma_1\gamma_2\gamma} w_{\gamma_1} 
w_{\gamma_2} w_{\gamma} \right) \left( {C}^{N_1 N_2 
N}_{\gamma_1\gamma_2\gamma'} w_{\gamma_1} w_{\gamma_2} 
w_{\gamma'} \right) = {1\over \sqrt{{\rm dim}_{N_1} \ldots 
{\rm dim}_{N_4} }}.
\label{ultima}
\end{equation}

%%%%%%%%%%%%%%%%%%%%%%%%%%%%%%%%%%%%%%%%%%%%%%%%%%
%%%%%%%%%%%%%%%%%%%%%%%%%%%%%%%%%%%%%%%%%%%%%%%%%%


\begin{thebibliography}{10}
 
\bibitem{Barrett:1998}
   J.~W.\ Barrett and L.\ Crane, J.\ Math.\ Phys.\ {\bf 39} 
   (1998) 3296.
 
\bibitem{Reisenberger:1997}
   M.~P.\ Reisenberger and C.\ Rovelli, Phys.\ Rev.\ {\bf 
   D56} (1997) 3490.  C.\ Rovelli, Nucl.\ Phys.\ {\bf B57} 
   (1997) 28.  C.\ Rovelli, gr-qc/9806121.
 
\bibitem{loop}
   C.\ Rovelli, ``Loop quantum gravity'', in {\em Living 
   Reviews in Relativity\/}, electronic journal, 
   http://www.livingreviews.org, gr-qc/9710008.
 
\bibitem{Baez}
   J.\ Baez, Class.\ Quant.\ Grav.\ {\bf 15} (1998) 1827.
 
\bibitem{spinnetwork}
   C.\ Rovelli and L.\ Smolin, Phys.\ Rev.\ {\bf D52} (1995) 
   5743.  J.\ Baez, Adv.\ Math.\ {\bf 117} (1996) 253.
 
\bibitem{Baez:1994}
   J.\ Baez, Strings, loops, knots and gauge fields, in {\em 
   Knots and Quantum Gravity} , ed J.\ Baez, Oxford 
   University Press 1994.
    
\bibitem{Reisenberger}
M. \ Reisenberger, ``Worldsheet formulations of gauge 
theories and gravity'', talk given at the 7th Marcel 
Grossmann Meeting Stanford, July 1994, gr-qc/9412035.
 
\bibitem{Iwasaki} J.\ Iwasaki, Journ Math Phys 36 (1995) 
6288.
 
\bibitem{Baez2}
J.~Baez, ``An introduction to spin foam models of BF theory 
and quantum gravity'', to appear in to appear in Geometry 
and Quantum Physics, eds.  Helmut Gausterer and Harald 
Grosse, Lecture Notes in Physics, Springer-Verlag, Berlin.  
{\it xxx-archive:}gr-qc/9905087

\bibitem{othersn}
   M.\ Reisenberger, gr-qc/9711052.  F.\ Markopoulou, 
   gr-qc/9704013.  F.\ Markopoulou, L.\ Smolin Phys Rev 
   D58:084032 (1998).  L.\ Smolin, hept-th/9001022, 
   hep-th/9808192.  R.\ De Pietri, {\it Canonical "loop" 
   quantum gravity and spin foam models}, to appear in the 
   proceedings of the XXIIIth Congress of the Italian 
   Society for General Relativity and Gravitational Physics 
   (SIGRAV), 1998, gr-qc/9903076.  J.~W.\ Barrett and L.\ 
   Crane, gr-qc/9904025.  J.\ Iwasaki, gr-qc/9903122.

\bibitem{altridue}
   J.~W.\ Barrett and R.\ Williams, gr-qc/9809032.
        
\bibitem{2d}
   F.\ David, Nucl.\ Phys.\ {\bf B257} (1985) 45.  J.\ 
   Ambjorn, B.\ Durhuus, J.\ Frolich, Nucl.\ Phys.\ {\bf 
   B257} (1985) 433.  V.\ A.\ Kazakov, I.\ K.\ Kostov, A.\ 
   A.\ Migdal, Phys.\ Lett.\ {\bf 157} (1985) 295.  D.\ V.\ 
   Boulatov, V.\ A.\ Kazakov, I.\ K.\ Kostov, A.\ A.\ 
   Migdal, Nucl.\ Phys.\ {\bf B275} (1986) 641.  M.\ 
   Douglas, S.\ Shenker, Nucl.\ Phys.\ {\bf B335}(1990) 635.  
   D.\ Gross, A.\ Migdal, Phys.\ Rev.\ Lett.\ {\bf 64} 
   (1990) 635; E.\ Brezin, V.\ A.\ Kazakov, Phys.\ Lett.\ 
   {\bf B236} (1990) 144.
 
\bibitem{Boulatov:1992}
   D.~V.\ Boulatov, Mod.\ Phys.\ Lett.\ {\bf A7} (1992) 
   1629.
 
\bibitem{Ponzano:1968}
  G.\ Ponzano and T.\ Regge, in {\em Spectroscopy and group 
  theoretical methods in Physics}, edited by F. Bloch 
  (North-Holland, Amsterdam, 1968).
 
\bibitem{Turaev-Viro}
  V.\ G.\ Turaev and O.\ Y.\ Viro, Topology {\bf 31} (1992) 
  865--902.  V.\ G.\ Turaev, {\em Quantum Invariants of 
  Knots and 3-manifolds} (de Gruyter, New York 1994).
 
\bibitem{Ooguri:1992b}
  H.\ Ooguri, Mod.\ Phys.\ Lett.\ {\bf A7} (1992) 2799.
 
\bibitem{Crane-Yetter}
  L.\ Crane and D.\ Yetter, ``A Categorical construction of 
  4-D topological quantum field theories'', in {\it Quantum 
  Topology}, L.\ Kaufmann and R.\ Baadhio eds., World 
  Scientific, Singapore 1993.  hep-th/9301062.
 
\bibitem{Crane}
  L.\ Crane, L.\ Kauffman and D.\ Yetter, J.\ Knot Theor.\ 
  Ramifications {\bf 6} (1997) 177--234, hep-th/9409167.
 
\bibitem{BF}
        G.\ Horowitz, Comm.\ Math.\ Phys.\ {\bf 125} (1989) 417.  
        M.\ Blau and G.\ Thompson, Phys.\ Lett.\ {\bf B228} 
        (1989) 64.  J.\ Baez, Lett.\ Math.\ Phys.\ {\bf 38} 
        (1996) 129.

\bibitem{DF}
   R.\ De Pietri and L.\ Freidel, Class.\ Quant.\ Grav.\ 
   {\bf 16} (1999) 2187, gr-qc/9804071.  M.\ Reisenberger, 
   Class.\ Quant.\ Grav.\ {\bf 16} (1999) 1357, 
   gr-qc/9804061.
 
\bibitem{Barbieri}
A.\ Barbieri, Nucl.\ Phys.\ {\bf B518} (1998) 714.  L.\ 
Freidel and K.\ Krasnov, Class.\ Quantum Grav.\ {\bf 16} 
(1999) 351.  J.\ Baez and J.\ Barrett, gr-qc/9903060.
 
\bibitem{FW} L.\ Freidel and K.\ Krasnov Adv.\ Theor.\ 
Math.\ Phys.\ {\bf 2} (1998) 1221.

\bibitem{Freidel:1999b}
   L.\ Freidel, K.\ Krasnov, and R.\ Puzio, hep-th/9901069.  
   L.\ Freidel and K.\ Krasnov, hep-th/9903192.
 
\bibitem{Peskin} M.E.\ Peskin, D.V.\ Schroeder, {\em An 
Introduction to Quantum Field Theory}, (Addison Wesley, 
Reading 1995).

\bibitem{Petronio}
R.\ Benedetti and C.\ Petronio, Manuscripta Mathematica {\bf 
88} (1995) 291-310.  C.\ Petronio, {\it Standard Spines and 
3-Manifold}, Collana Tesi di Perfezionamento, Scuola Normale 
Superiore, Pisa, 1995.  See also R.\ Benedetti and C.\ 
Petronio, {\it Branched Standard Spine of 3-Manifolds}, 
Lecture Notes in Mathematics N. 1653 (Springer Verlag, 
Berlin, 1997).
        
\bibitem{Carfora}
   J.\ Ambjorn, M.\ Carfora, A.\ Marzuoli, {\it The Geometry 
   of Dynamical Triangulations}, Lecture Notes in Physics, 
   (Springer-Verlag, Berlin 1997).  J.\ Ambjorn, M.\ 
   Carfora, D. Gabrielli and A.\ Marzuoli, Nucl.\ Phys.\ 
   {\bf B542} (1999) 349.
 
\bibitem{Roberto}
        R.\ De Pietri, ``Critical behavior in the convergence of 
        sums over triangulations'' unpublished.
 
\bibitem{Mike}
        M.\ Reisenberger, J.\ Math.\ Phys.\ {\bf 40} (1999) 
        2046.
 
\bibitem{VolLoop}
        C.\ Rovelli and L.\ Smolin, Nucl.  Phys.  {\bf B442} 
        (1995) 593, Erratum-ibid.  {\bf B456} (1995) 753.  R.\ 
        De Pietri and C.\ Rovelli, Phys.\ Rev.\ {\bf D54} (1996) 
        2664.
 
\bibitem{Vilenkin:1993b}
        N.~J. Vilenkin and A.~U. Klimyk, {\em Representation of 
        Lie Groups and Special Functions} (Klewer Academic 
        Publisher, Dordrecht, The Netherland, 1993), Vol.~2, 
        volume 2: Class I Representations, Special Functions, 
        and Integral Transforms.
 
\end{thebibliography}
\end{document}